\documentclass[twoside,slac_one]{revtex4}
\usepackage{graphicx}
\usepackage{fancyhdr}
\usepackage{amsmath} % American Mathematics Society standards
\usepackage{bm}% bold math
\usepackage{amsxtra}
\usepackage{amssymb}
\usepackage{amsthm}
\usepackage{latexsym}
\usepackage{lscape}
\usepackage{xspace}

\begin{document}

\input{macros}  

\title{Rates of Jets Produced in Association with W and Z Bosons}


\author{K.S. Grogg on behalf of the CMS Collaboration}
\affiliation{Department of Physics, University of Wisconsin, Madison, WI, USA}

\begin{abstract}
Presented here is a study of jets produced in association with vector bosons production in pp collisions at $\sqrt{s} = 7$~TeV using the full CMS 2010 data set, corresponding to an integrated luminosity of $36\pm1.4~pb^{-1}$. The transverse energy distribution of the reconstructed leading jets is measured and compared to theoretical expectations. The jet multiplicity distributions are  corrected for efficiency and unfolded. The ratios of multiplicities, $\sigma(\mathrm{V}+\ge n~{\mathrm{jets}})/
\sigma(\mathrm{V}+\ge (n-1)~{\mathrm{jets}})$ 
and $\sigma (\mathrm{V}+\ge n~{\mathrm{jets}})/\sigma(\mathrm{V})$ where n stands for number of jets, are also presented along with the first test of the Berends-Giele scaling at $\sqrt{s} = 7$~TeV. 

\end{abstract}

\maketitle

\thispagestyle{fancy}


%%%%%%%%%%%%%%%%%%%%%%%%%%%%%%%%%%
\section{Introduction}
%The study of jets produced in association with W and Z vector bosons
%(denoted \PV) provides a rigorous test of perturbative QCD calculations. 
The production of hadronic jets in association with \W and \Z vector bosons (denoted V+jets) provides the means for a rigorous study of perturbative QCD.
Because the production of vector bosons with jets constitutes a significant source of 
background in searches for new physics and for studies of the top
quark, a precise measurement of the \vnjets cross section and an
understanding of the kinematics is essential.
\par
This proceedings presents results obtained with the 2010 data sample of
the CMS experiment at the Large Hadron Collider (LHC) at CERN, based
on $36.1\pm 1.4~\pbi$ of integrated luminosity collected in
proton-proton collisions at $\sqrt{s}$ = 7 TeV.
To reduce theoretical and systematic uncertainties, we measure the \vnjets cross sections relative to
the inclusive
W and Z cross sections.  We also measure the cross section ratios
$\sigma(\PV +\ge n~{\mathrm{jets}})/\sigma(\PV +\ge (n-1)~{\mathrm{jets}})$ from which we are able to test Berends-Giele
scaling~\cite{BGscaling}.
The complete V+jets analysis is reported in~\cite{VJets}.

%%%%%%%%%%%%%%%%%%%%%%%%%%%%%%%%%%
\section{Samples and Reconstruction}

Monte Carlo (MC) simulation samples are used both for comparison to data and to unfold the jet multiplicity distributions.
Simulated events with a \W or a \Z boson are generated with the
\MADGRAPH~\cite{Maltoni:2002qb} event generator, producing parton-level events with a vector boson and up to
four jets on the basis of a matrix-element calculation. 
\MADGRAPH is interfaced to the \PYTHIA~\cite{Sjostrand:2006za} program for parton shower simulation.  
Top pair (\ttbar) and single top processes
are generated with \MADGRAPH also. Multijet and $\gamma$+jets processes are generated with
\PYTHIA alone. 
For comparison to the data distributions,
the simulation samples are normalized to NNLO or NLO
cross sections and scaled to the luminosity. 
The PYTHIA parameters for the underlying
event are set to the ``Z2'' tune, a modification of the ``Z1'' tune described in~\cite{Field:2010bc}.
Comparisons are also made to the ``D6T'' tune~\cite{Skands:2010ak}.  Minimum-bias events are superimposed on the simulated events to represent ``pile--up'' found in data from multiple proton interactions in a single bunch crossing.

Muons are reconstructed using both the silicon tracker and muon chambers.
Identification based on compatibility sub-detector measurements is used to assure quality muons with \PT resolution of about 1-2\%.
Electron candidates are produced by matching tracks to superclusters constructed from EM calorimeter (ECAL) energy deposits with an \ET resolution of about 1\%.

A particle flow (PF) algorithm is used to reconstruct both the missing transverse energy (\MET) and the jets in the event.
The PF algorithm creates a complete event description by collecting information from all of the sub--detectors and linking it together. Objects are initially formed into the categories of muons, electrons, photons, charged hadrons, and neutral hadrons. From these elements the jets and missing transverse energy, \MET, are reconstructed.

The \MET is reconstructed as the opposite of the sum of the transverse momentum of all of the PF particles.  
Jets are reconstructed from PF objects by means of the anti-$k_T$ algorithm~\cite{ref:antikt} with a size parameter of $R = 0.5$.  
Jet energy corrections (JEC) are applied to account for the jet energy response as a function of $\eta$
and \pt and corrections are made to the jet energy for the effect of pile--up.

%%%%%%%%%%%%%%%%%%%%%%%%%%%%%%%%%%
\section{Signal Selection}
Signal selection begins with the identification of a ``leading 
lepton'', either an electron or a muon.  
The bulk of the lepton selection follows the standard established by the
measurement of the inclusive \W and \Z cross sections~\cite{Khachatryan:2010xn}.

For electron candidates we require \PT$>$~20~GeV and that the ECAL cluster lies in
the fiducial region of $|\eta| < 2.5$ while excluding the region $1.4442 < |\eta| <
1.566$ in order to reject electrons close to the
barrel/endcap transition where cables and services reduce detectability.
A series of quality requirements including 
identification, isolation, and conversion rejection are then applied to the electron.
For the leading electron, the values of the different quality requirements are chosen such that they
correspond to an electron efficiency of about 80\% as evaluated with a \MADGRAPH+ \PYTHIA simulated sample.
\par
If there is a second electron of \PT$> 10$~GeV and it is within the ECAL fiducial volume, passes a looser set of quality cuts (corresponding at an efficiency of about 95\%), and forms an invariant mass with the leading electron between $60~\GeV$ and
$120~\GeV$, then the event is placed in the \zjetsb~sample. Otherwise, the event is assigned to the \wjets sample. 
Events with a muon with \pt $ > 15~\GeV$ and $|\eta| < 2.4$ 
are then rejected from the \wjets sample to reduce \ttbar~contamination.
\par
The muon selection starts by requiring the presence of an isolated muon in the region 
 $|\eta| < 2.1$ with \pt~$>~20$~\GeV passing the requirements described in~\cite{Khachatryan:2010xn} along with 
a transverse impact parameter $|d_{xy}| < 2$~mm to suppress cosmic--ray muon background.  
A requirement that the combined activity of the tracker and calorimeters around the muon is less than 0.15 relative to the muon \PT  results in quality muons and background suppression.
If there is (is not) a
second muon of \pt$ > 10$~\GeV accepted in the range $|\eta| < 2.5$ such that the dimuon invariant mass lies within
the region 60~\GeV~to~120~\GeV, then the event is assigned to the \zjetsb~(\wjets) sample.
\par
For both \wjets samples, the transverse mass, \MT, is constructed from the lepton and \MET,
$\MT~=~\sqrt{2\PT\MET (1-\cos\Delta\phi)}$
where $\Delta\phi$  is the angle in the $xy$-plane. To avoid a region at low~$\MT$ containing essentially no signal we require that
\MT~$> 20$~\GeV.

\section{Jet Rates}

\begin{figure}%[h!]
  \centering
\includegraphics[width=0.4\textwidth]{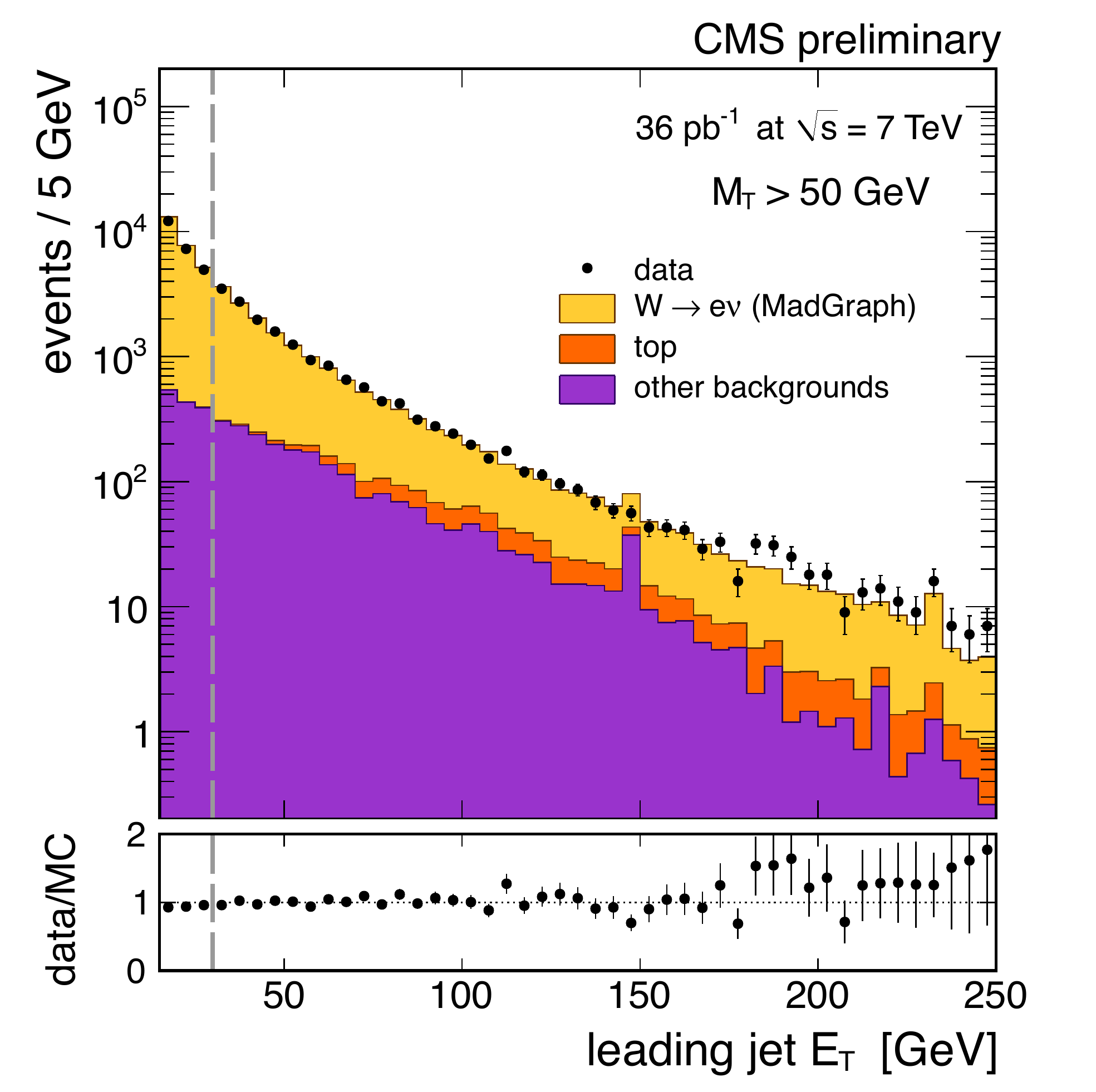}
\includegraphics[width=0.4\textwidth]{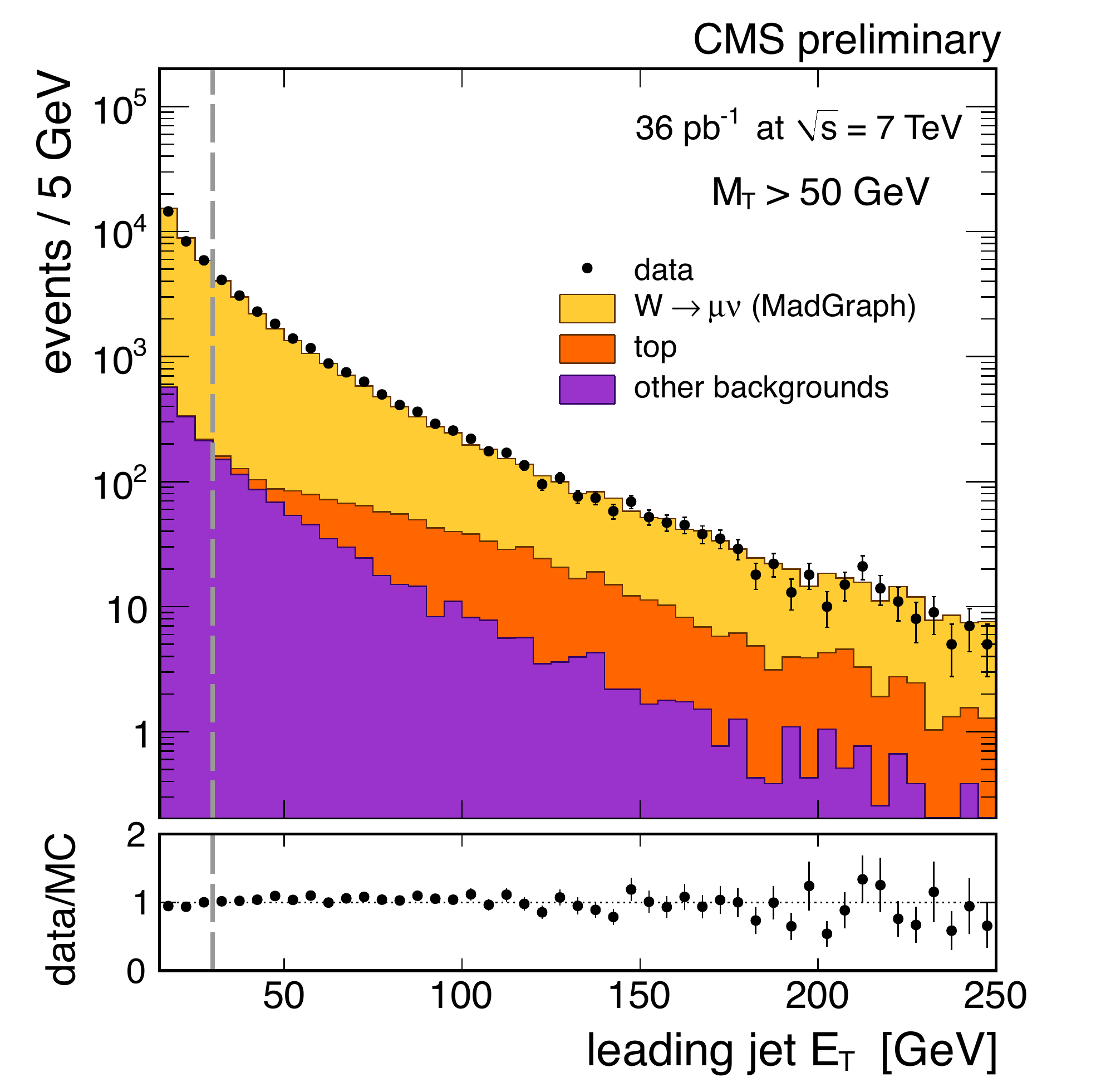}
  \caption{
Distributions of the uncorrected \PT for the leading jet
in the $W + 1$~jet sample for the electron channel (left)
and for the muon channel (right).
The ratio between the data and the simulation is also shown.
The line at $\PT = 30$~\GeV~corresponds to the threshold
imposed for counting jets.
\label{fig:leadjetPt_W}}

\centering
\includegraphics[width=0.4\textwidth]{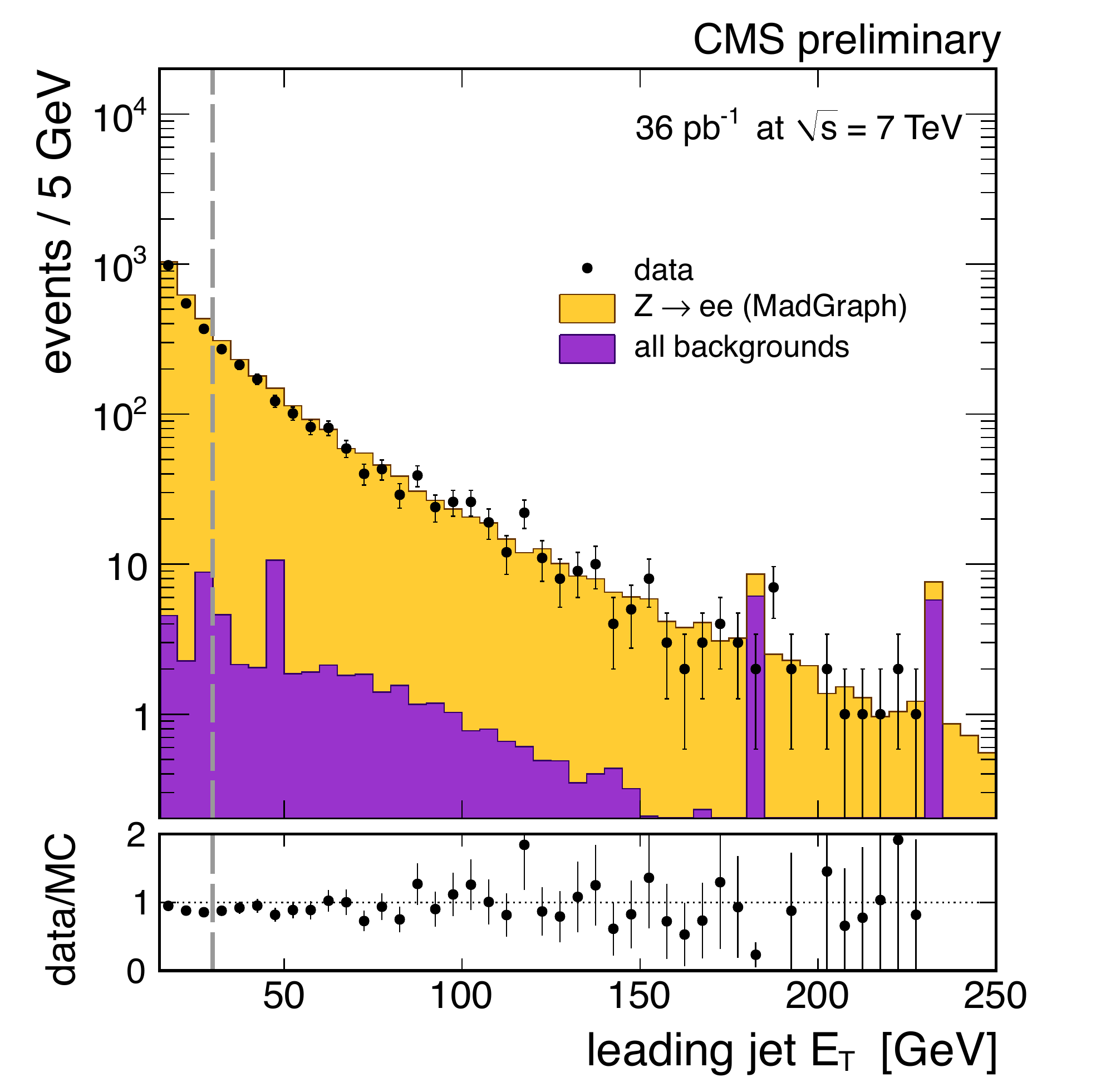}
\includegraphics[width=0.4\textwidth]{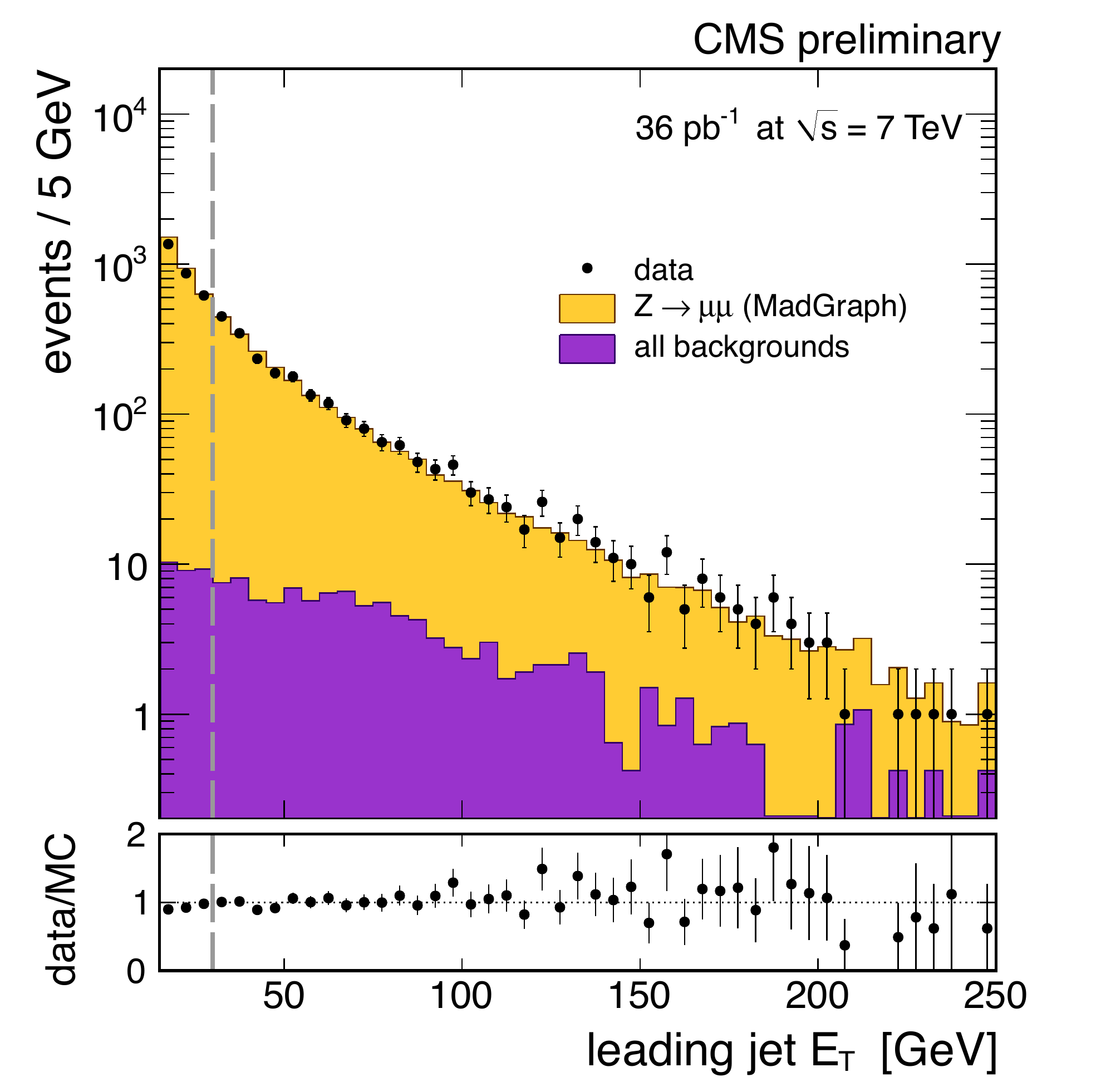}
  \caption{Distributions of the uncorrected \PT for the leading jet
in the $Z + 1$~jet sample for the electron channel (left)
and for the muon channel (right).
The ratio between the data and the simulation is also shown.
The line at $\PT = 30$~\GeV~corresponds to the threshold
imposed for counting jets. 
\label{fig:leadjetPt_Z}}
\end{figure}

Jets must first satisfy identification criteria to eliminate jets originating from
noise in the calorimeter. We require that the jets fall within the tracker acceptance of $|\eta| < 2.4$.
The observed transverse momentum distributions for the leading jet
are shown in Figs.~\ref{fig:leadjetPt_W}~and~\ref{fig:leadjetPt_Z}.
The data is in good agreement with the \MADGRAPH predictions 
normalized to the NNLO cross sections. 
For the \W sample, we have required $\MT > 50$~\GeV~in order to reduce backgrounds.

\begin{figure}%[h!]
  \centering
\includegraphics[width=0.45\textwidth]{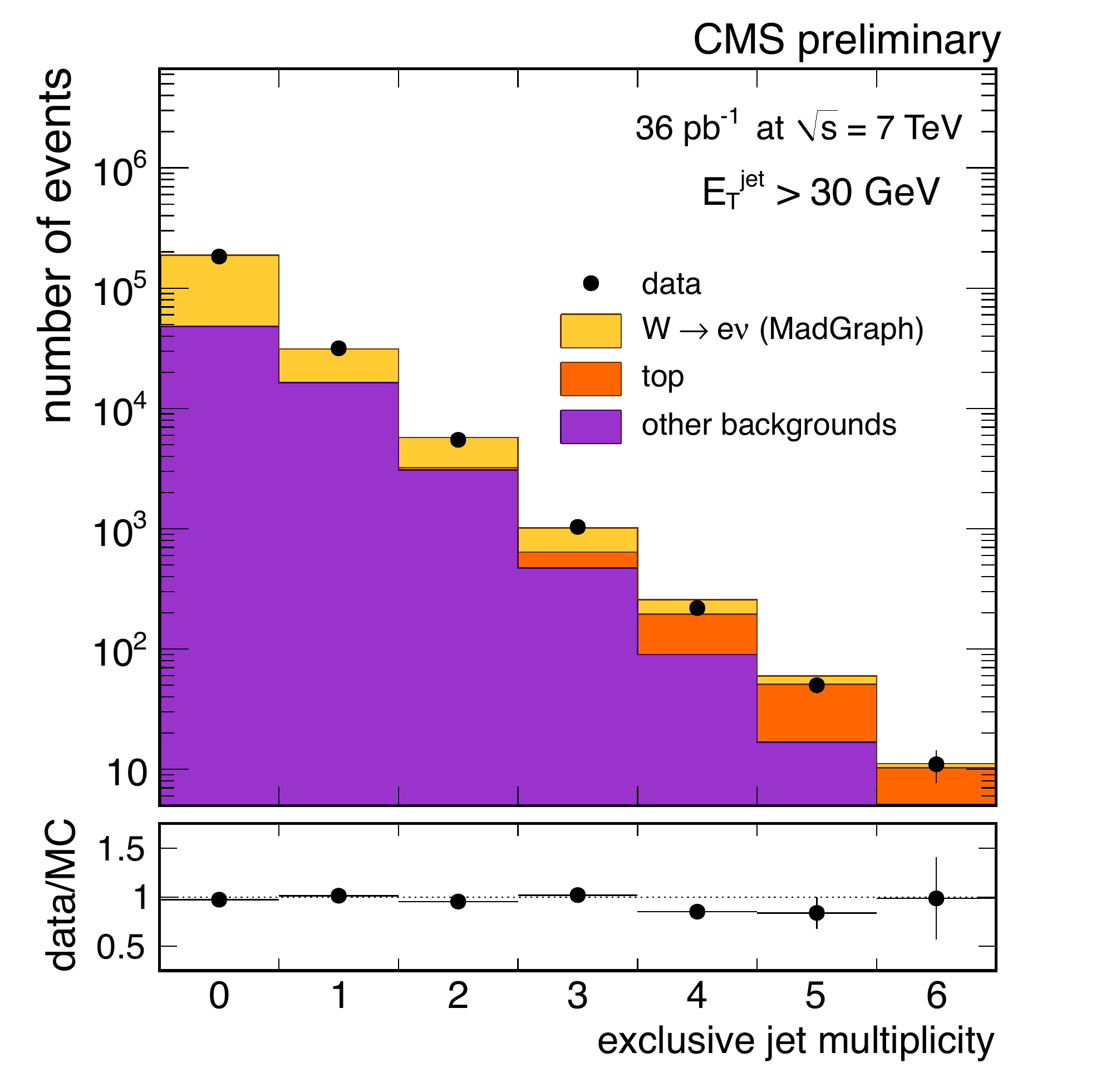}
\includegraphics[width=0.45\textwidth]{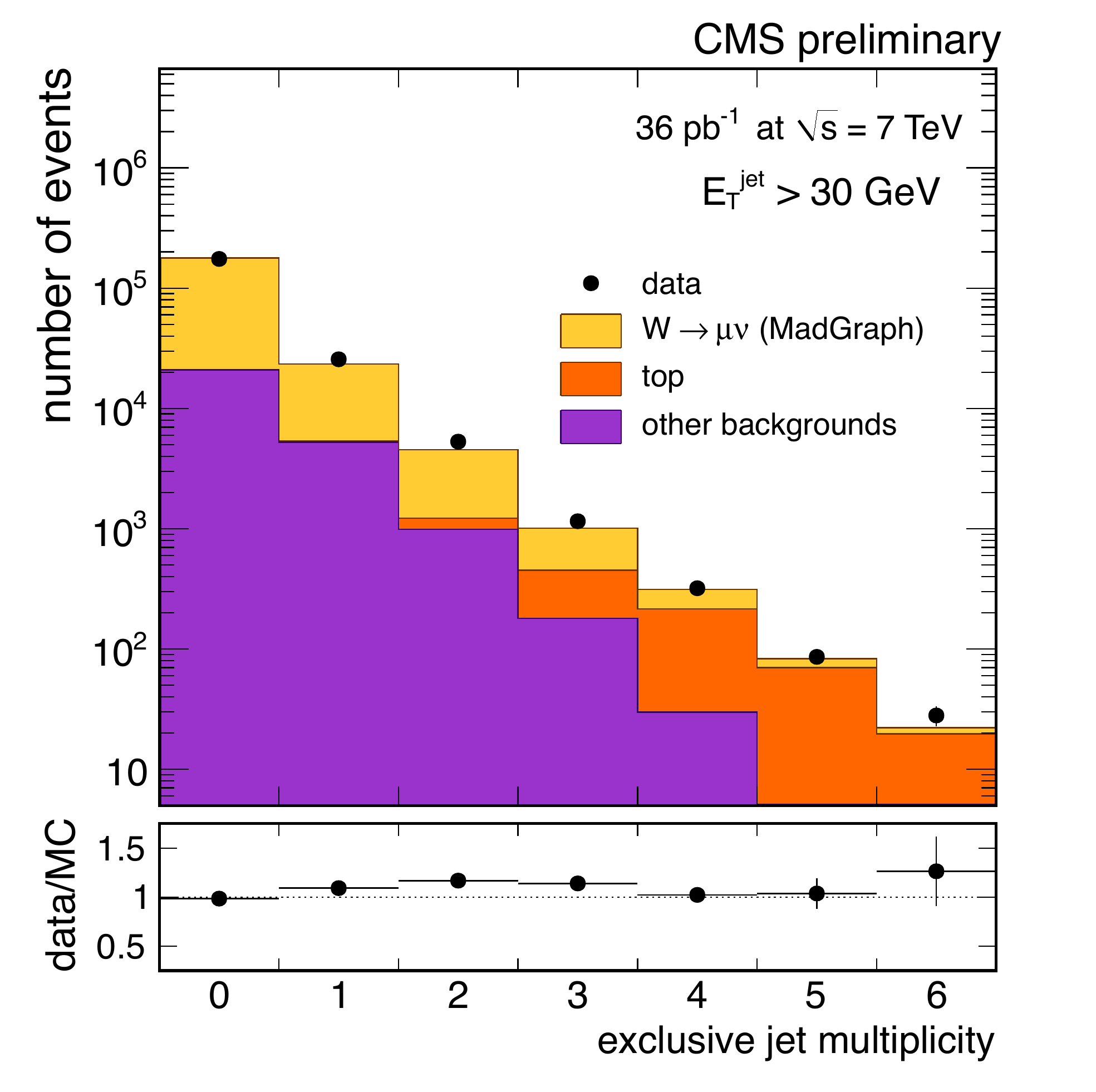}
  \caption{Exclusive number of reconstructed jets in events
    with $\Wen$ (left) and $\Wmn$ (right).  The histograms
    represent the expectations based on simulated events.
\label{fig:Wrawrates}}
\end{figure}

\begin{figure}%[h!]
  \centering
\includegraphics[width=0.45\textwidth]{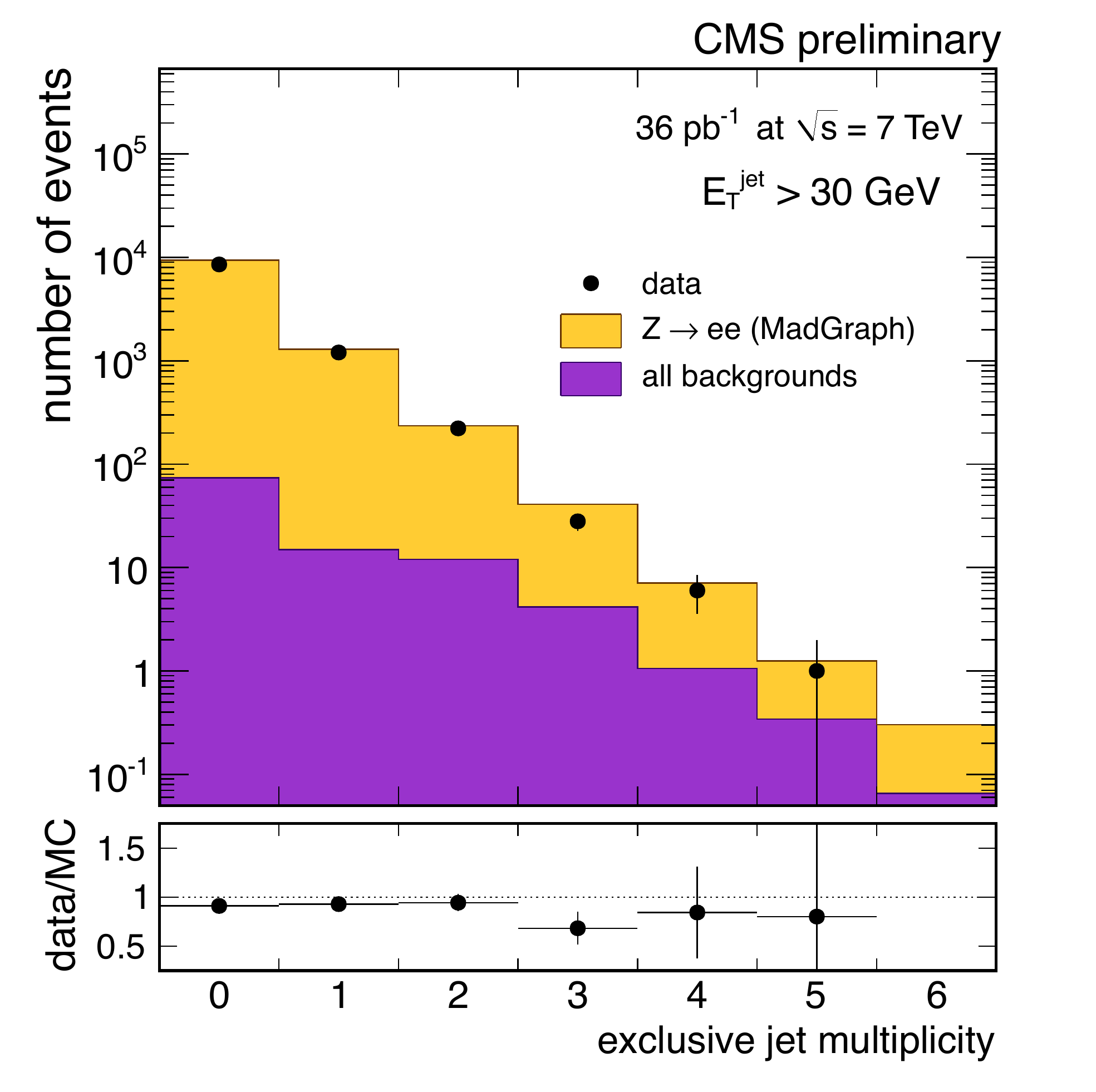}
\includegraphics[width=0.45\textwidth]{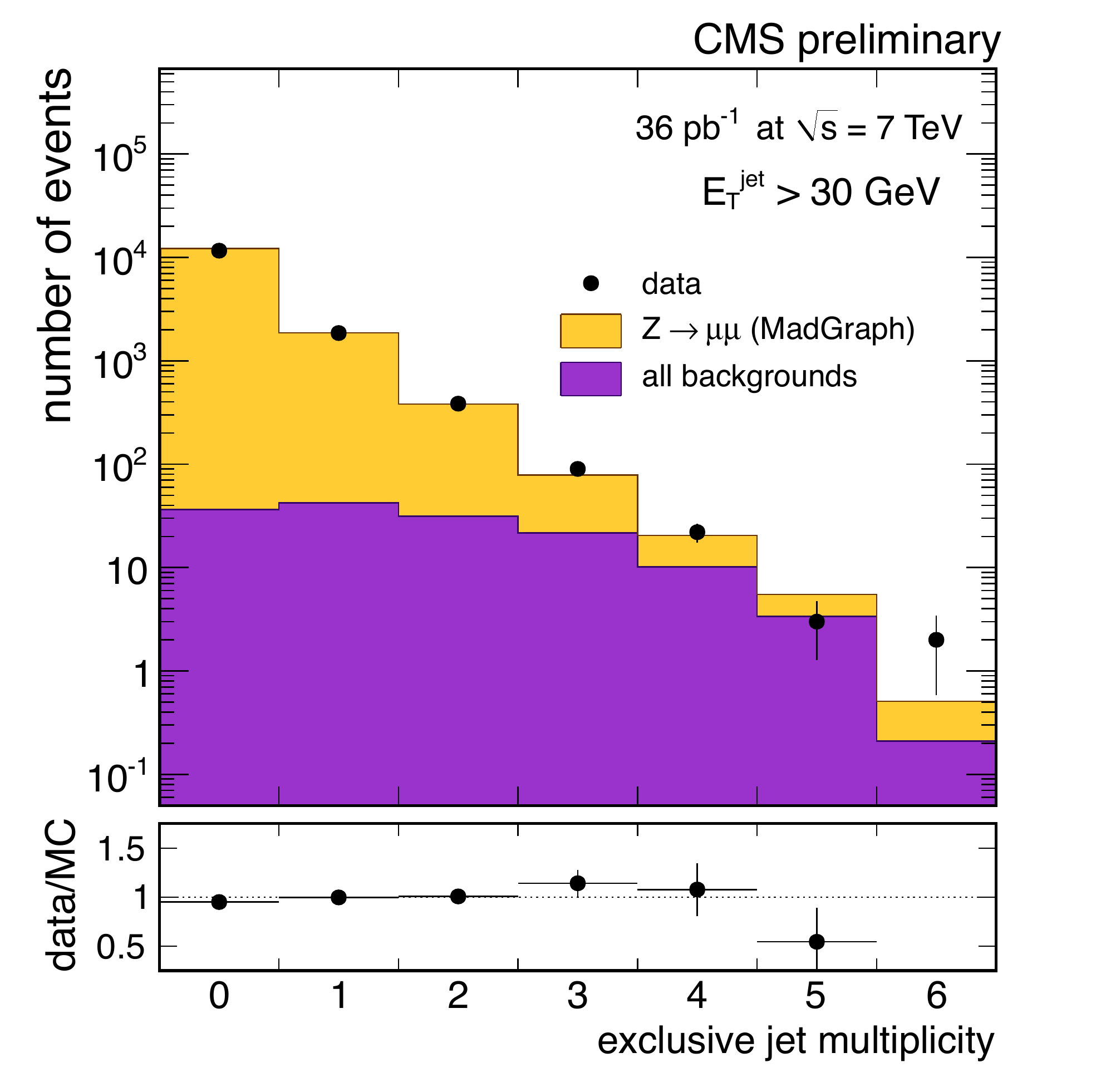}
  \caption{Exclusive number
 of reconstructed jets in events
    with $\Zee$ (left) and $\Zmm$ (right).  The histograms
    represent the expectations based on simulated events.
\label{fig:Zrawrates}}
\end{figure}

Events are assigned to exclusive bins of jet
multiplicity by counting the number of jets in the event with \pt~$> 30$~\GeV.
The observed distributions of the exclusive 
numbers of reconstructed jets in the \W\ and \Z samples are 
shown in Figs.~\ref{fig:Wrawrates} and~\ref{fig:Zrawrates}, 
respectively.   The distributions from simulation are also shown, with overall good
agreement.

One of the most important backgrounds in the \W sample
comes from \ttbar~events. These events contain two $b$-quark jets.
Jets are $b$-tagged with a tagging algorithm that requires at least two tracks in the
jet with a significance on the transverse impact parameter greater
than 3.3.  This choice of cut results in a $b$-tagging efficiency of about 
62\% and a mis--tagging rate of about 2.9\%~\cite{btv-10-001}.  
The number of $b$-tagged jets, $\NBJET$, is used in the fitting method to separate \W from top events.

\section{Acceptance and Efficiency}
\label{sec:efficiency}
\par

In order to avoid model--dependent results, we quote all results within the lepton and jet acceptance, and only correct for efficiency of the selection.
The efficiencies for lepton reconstruction, identification, isolation and 
trigger are obtained by a tag-and-probe 
method performed on \zjets~data.
The tag--and--probe sample for the measurement of a given efficiency contains events selected with two lepton 
candidates of invariant mass in the range [60-120]~GeV. 
One lepton candidate, called the  ``tag'', satisfies all  selection
requirements. The other lepton candidate, called the ``probe'', 
is selected with criteria that depend on which efficiency is being examined. 
The signal yields are obtained both for events in which the probe lepton passes or in which it
fails the selection criteria considered. 

Fits are performed to the invariant--mass distributions of the pass and fail subsamples to extract the Z signal events.
The measured efficiency is calculated from the relative level of signal in the pass and fail subsamples.
The lepton selection efficiency is the product of the reconstruction efficiency, the identification and isolation efficiency, and the trigger efficiency. 
Each of these efficiencies is calculated as a function of the jet multiplicity in the event. 
The ratio of tag--and--probe results for the \zjets data sample are combined with the full efficiency estimated from the W+jets and Z+jets selection in simulation.
\par
For electrons we find that the efficiencies are roughly 70\% (60\%) for the
\wjets (\zjets) signal events with variations of a few percent 
across different jet multiplicity bins.

\par
For muons, the efficiencies are measured as a 
function of \PT and $\eta$ in the highest statistics bins ($n = 0$ and $n = 1$).
Due to the isolation requirement, the efficiencies also exhibit 
a significant dependence on the observed jet multiplicity. 
Since the statistical precision in the bins with $n > 1$ is 
insufficient, the efficiencies for these bins are extrapolated 
by using the \PT and $\eta$ shape of the $n = 1$ bin.
We find an average efficiency close to 82\% for the leading \pt muon and of
above 90\% for the second leading muon.

%%%%%%%%%%%%%%%%%%%%%%%%%%%%%%%%%%
\section{Signal Exraction}
\label{sec:Extraction}

\par
The signal yield is extracted using an extended likelihood
fit to the invariant mass, $\MLL$, for the \zjetsb sample and to $\MT$ for the
\wjets sample.  
The fitting functions are parameterized on simulation and as many parameters as possible are allowed to vary in the fit.
\par
For the \Z event samples, the main
background processes, dominated by \ttbar~and~\wjets, are
small and do not produce a peak in the $\MLL$ distribution,
so the $\MLL$ distribution can be split to two components, one
for the signal and one for all background
processes.
\par
For the \W sample, background contributions are divided
into two components, one which exhibits a peaking structure
in $\MT$, dominated by \ttbar, and another which does not,
dominated by QCD multi-jet events.  We perform a two-dimensional
fit to the $\MT$ distribution and the number of $b$-jets, $\NBJET$.
The $\MT$ distribution distinguishes
the signal from the non-peaking backgrounds, while $\NBJET$
distinguishes the signal and the other backgrounds from~$\ttbar$. The likelihood fit is built on the assumption that the signal has no b-jets. This implies that a component of \W~produced in association with heavy flavor jets is counted as background. Considering the statistical precision of the measurement, this assumption has negligible effects on the \wjets cross section calculation.
\par
\begin{figure}%[h!]                                                                                                                                     
  \centering                                                                                                                                     
\includegraphics[width=0.4\textwidth]{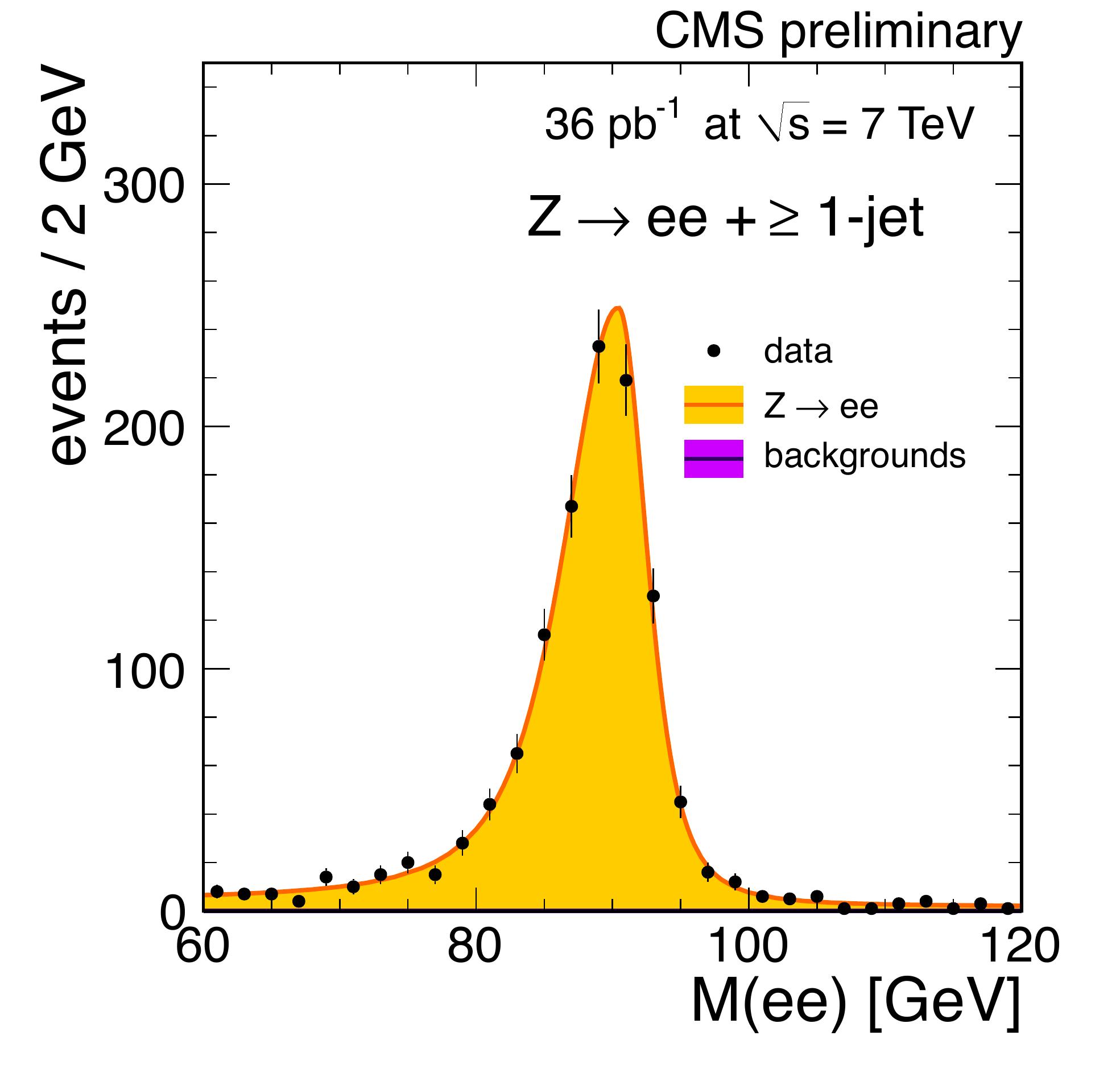}                                                                                                                                                                                                                                                                                    
\includegraphics[width=0.4\textwidth]{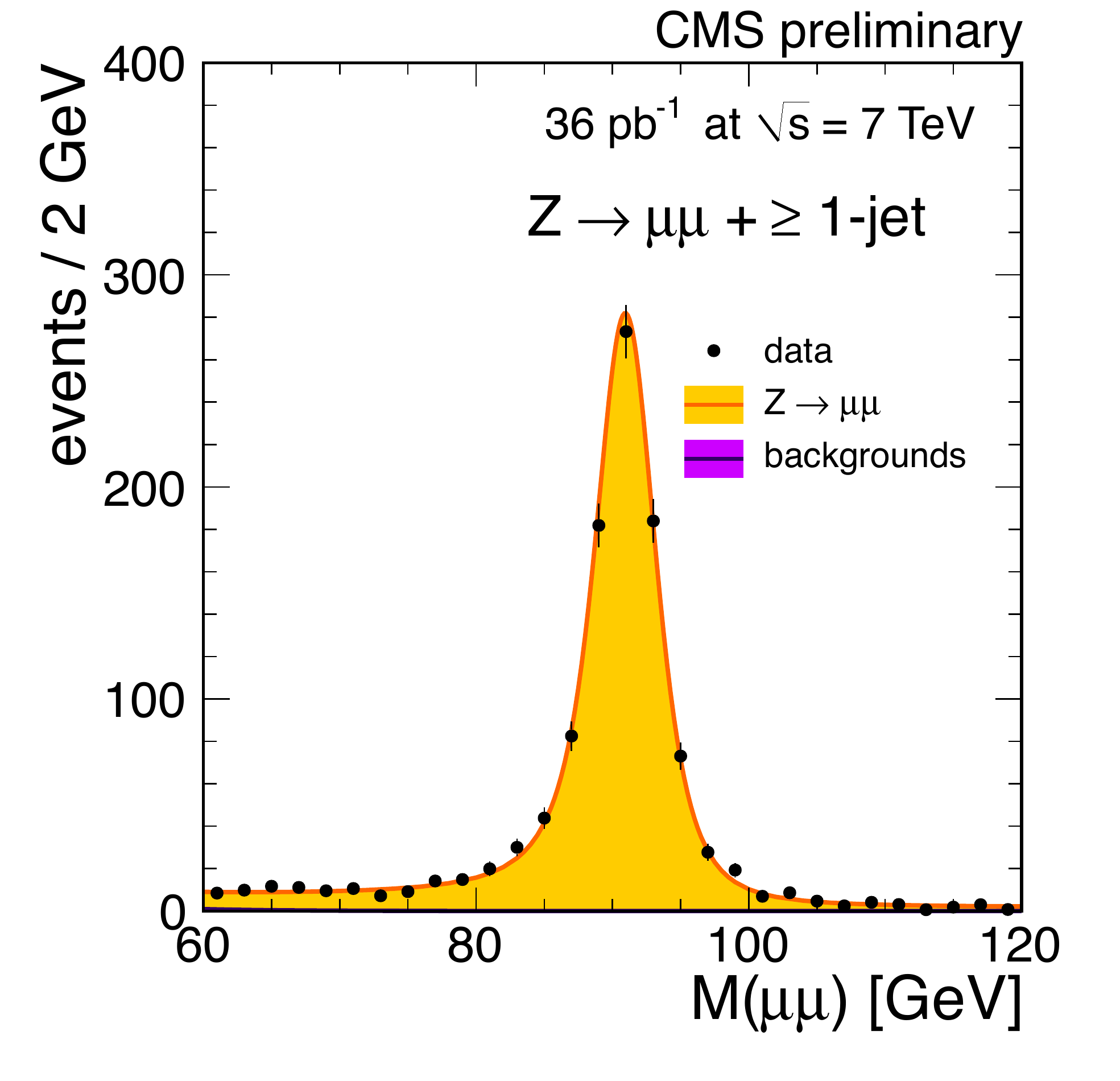}                                                                                                                                                     
  \caption{Di--lepton mass fit for the $Z+1$~jet samples, in the electron
channel (left) and the muon channel (right). The background is very
low, rendering hardly visible in the figure.
    \label{fig:zllfits}}
\end{figure}
\begin{figure}%[h!]                                                                                                                                     
  \centering                                                                                                                                      
\includegraphics[width=0.4\textwidth]{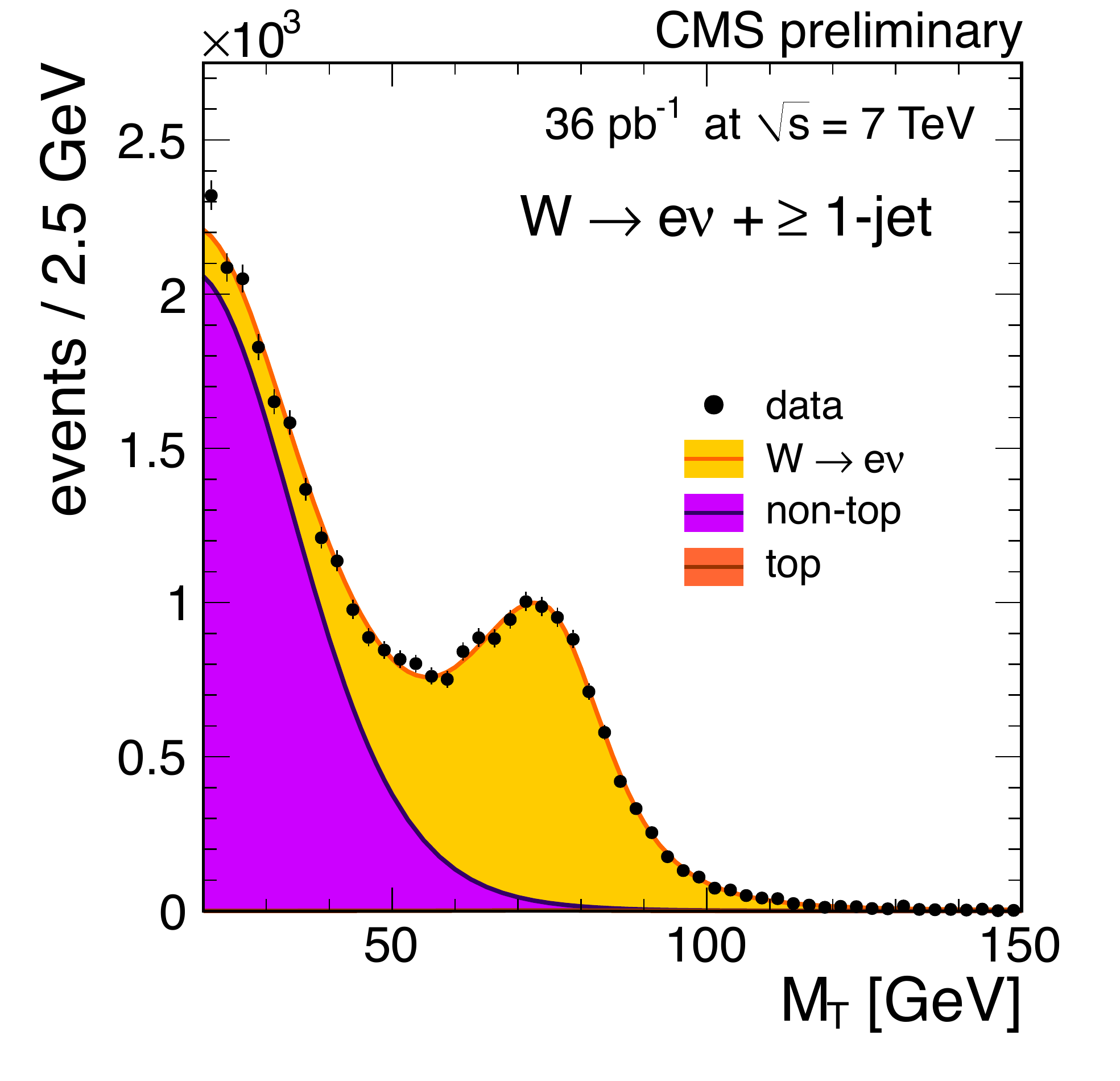}                                                                                                  
\includegraphics[width=0.4\textwidth]{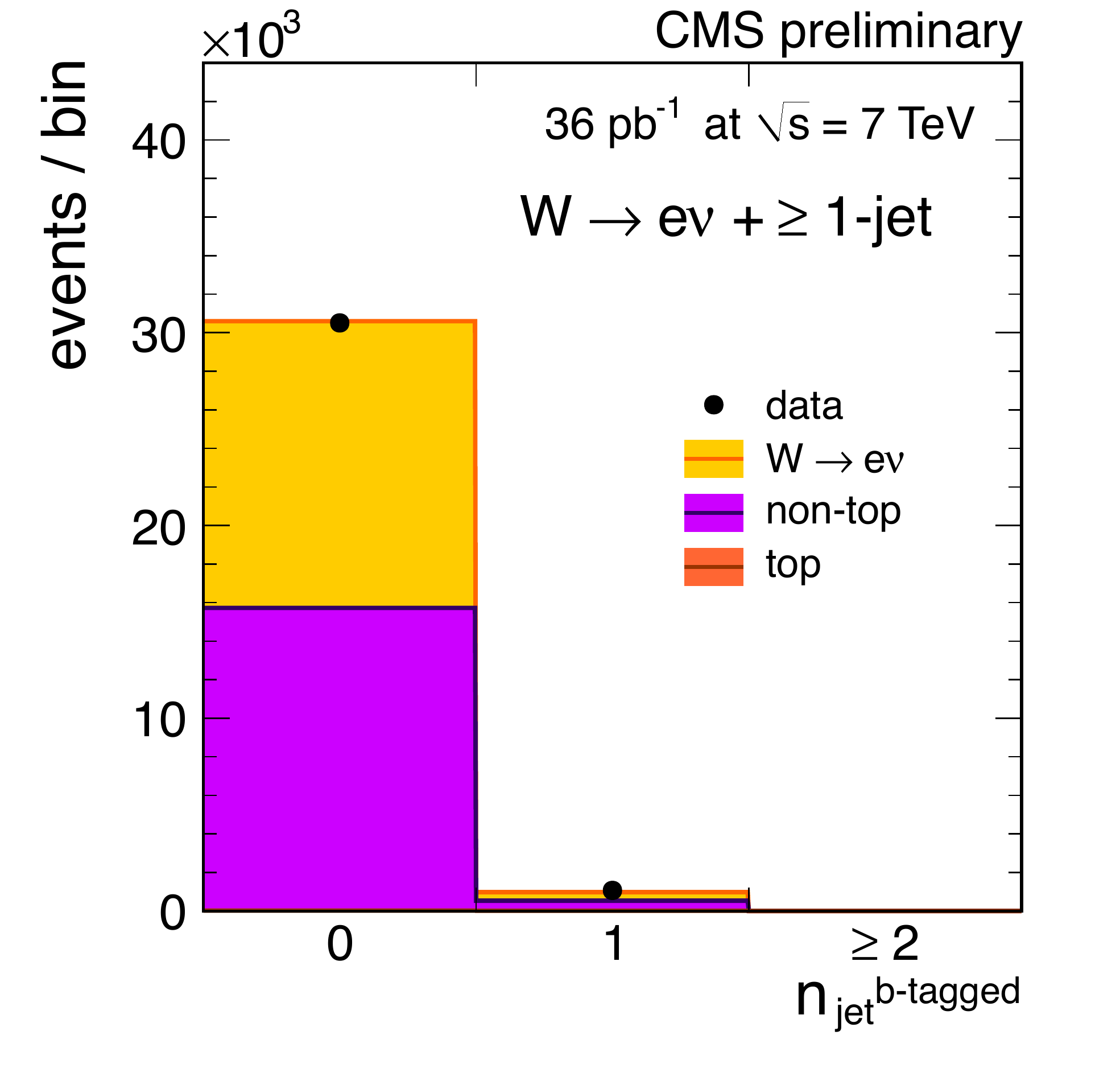}                                                                                          
  \caption{Fit results for the $W(e\nu)+ n$~jet sample with $n =1$.  On the left is
the \MT\ projection, and on the right $\NBJET$.
    \label{fig:wenufits}
    }
\end{figure}
\begin{figure}%[h!]                                                                                                                                     
  \centering                                                                                                
\includegraphics[width=0.4\textwidth]{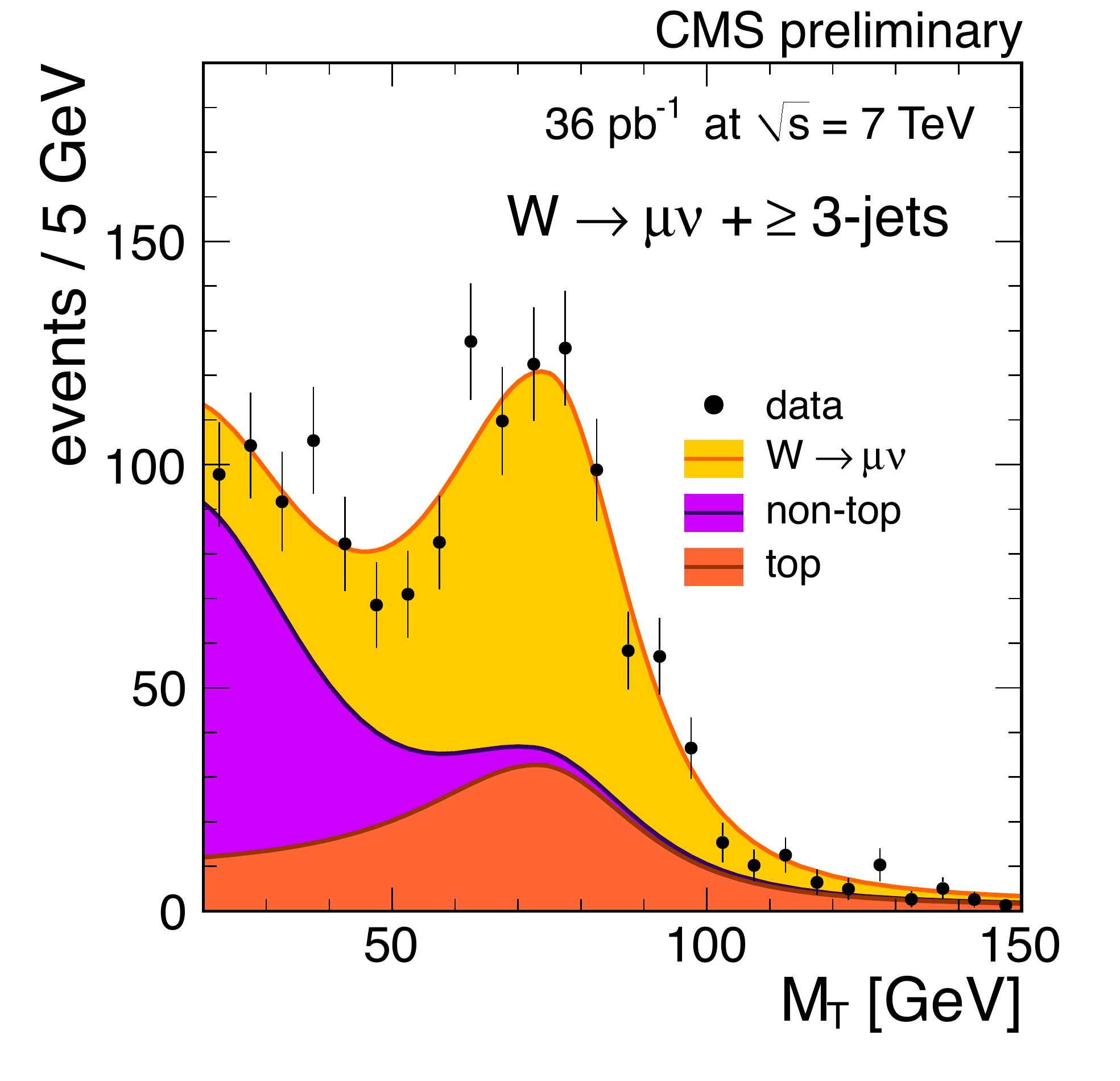}                                                                                           
\includegraphics[width=0.4\textwidth]{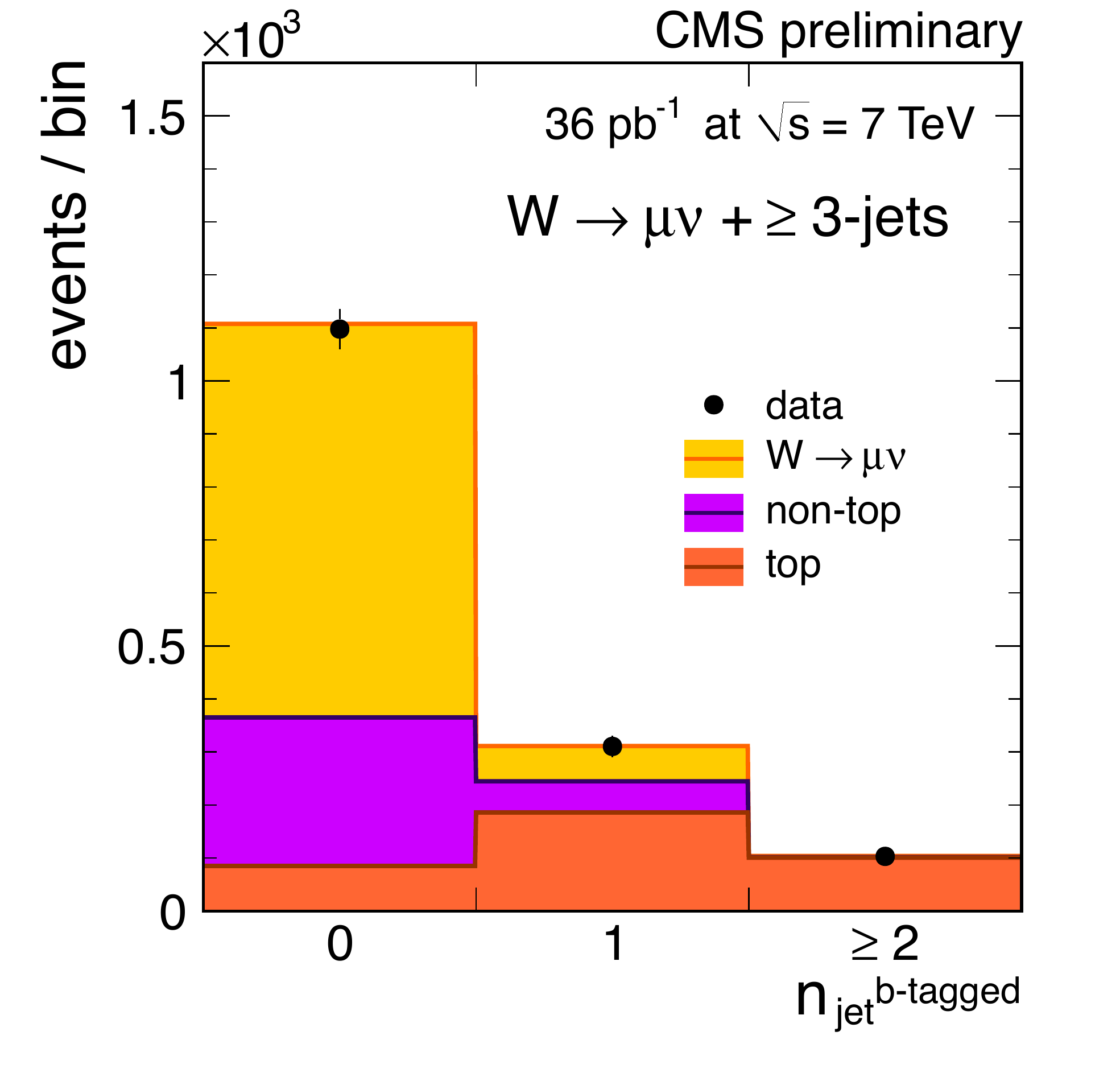}                                                                                          
  \caption{Fit results for the $W(\mu\nu)+ n$~jet sample with $n =3$.  On the left is
the \MT\ projection, and on the right $\NBJET$.
    \label{fig:wmunufits}
    }
\end{figure}

The fits are done in
exclusive jet multiplicity bins for $n \le 3$ and inclusively for the last
bin of jet multiplicity, {\it i.e.} $n \ge 4$.
Examples of fits for $Z + 1$~jet are shown in Figure~\ref{fig:zllfits}.
Figures~\ref{fig:wenufits} and~\ref{fig:wmunufits} show
fits in $\MT$ and $\NBJET$ projections for $W + n$~jets~(n=1 and n=3).
The presence of the top background
is evident comparing the $n = 1$ and $n = 3$ exclusive
multiplicity bins.

\par
In the electron channel, exclusive \vjets\ rates
are corrected for electron efficiencies as
discussed in Section~\ref{sec:efficiency}.
In the muon channel, efficiencies depend on the lepton
\PT and $\eta$ and on the jet multiplicity.
To account for these variations, every
event is assigned a weight and the fit is performed
to a weighted distribution.

A second fit is performed in order to test Berends-Giele scaling
and measure the associated parameters.  Events are
assigned to exclusive jet multiplicity bins and the yields are fit with the assumption
that they conform to a scaling function:
\begin{equation}
C_n = \frac{\sigma_n}{\sigma_{n+1}}
\label{eq:scaling}
\end{equation}
where $\sigma_n = \sigma(\PV+n~{\mathrm{jets}})$.
To first order one expects $C_n = \alpha$,
where the constant $\alpha$ is proportional to the inverse of the strong
coupling constant, $\alpha_S^{-1}$. Phase space effects can violate this simple proportionality,
so we introduce a second parameter, $\beta$,
to allow for a deviation from a simple constant scaling law:
$  C_n=\alpha + \beta \, n .$
Due to the different kinematics of the $n = 0$
sample, the scaling expressed in Eq.~(\ref{eq:scaling})
is not expected to hold, so we do not include the $n = 0$ sample
in the fit.

%%%%%%%%%%%%%%%%%%%%%%%%%%%%%%%%%%
\section{Unfolding}
\label{sec:unfolding}
\par
In order to estimate the scaling rule of jets at the 
{\it particle level}, we apply an unfolding procedure that 
removes the effects of jet energy resolution and
reconstruction efficiency.  A migration matrix, which
relates a number $n'$ of produced jets at particle level
to an observed number $n$ of reconstructed jets, is
derived from simulated samples of \zjetsb and \wjets 
with leptons and jets within their acceptances.
\par
We employ two well-known unfolding methods.  The base line 
method is the ``singular value decomposition'' (SVD)
method~\cite{Hocker:1995kb}.  As a cross check, we apply
the iterative or ``Bayesian'' method~\cite{D'Agostini:1994zf}.
Both algorithms require a regularization parameter, chosen to be $k_{SVD}=5$ and $k_{Bayes}=4$, to prevent 
the statistical fluctuations in the data from appearing as
structure in the unfolded distribution.   

%%%%%%%%%%%%%%%%%%%%%%%%%%%%%%%%%%
\section{Systematic Uncertainties} 

\renewcommand{\textfraction}{0}

One of the main sources of systematic uncertainties in the
W/Z+jets measurements is the jet energy scale (JES), which affects the jet counting.
The effect of jet energy uncertainties is evaluated on the jet multiplicity distribution using
simulations. Compatible results have been found in all
the channels, for both W and Z events. 
The pile-up subtraction was also tested comparing the jet multiplicity
in two simulated signal samples, one without pile--up, and one with pile--up
plus pile--up subtraction applied. The difference is found to be below 5\%. 

While the systematic uncertainty in the jet counting is correlated among the different jet
multiplicities, all other uncertainties, such as from efficiency and fits, are uncorrelated between jet multiplicities.
All statistical and both types of systematic uncertainties are propagated in the unfolding procedure. 
Finally, to estimate uncertainties in the unfolding procedure itself,
we calculated the difference in unfolding using the Bayes algorithm versus the SVD algorithm, and using two different simulations, \MADGRAPH and \PYTHIA,
for the unfolding matrix, and two different tunes, Z2 and D6T for the
unfolding matrix.   
The resulting uncertainties are shown with the final results in
the next section.

\section{Results and Conclusions}

\begin{figure}%[h!]
  \centering
  \includegraphics[width=0.4\textwidth]{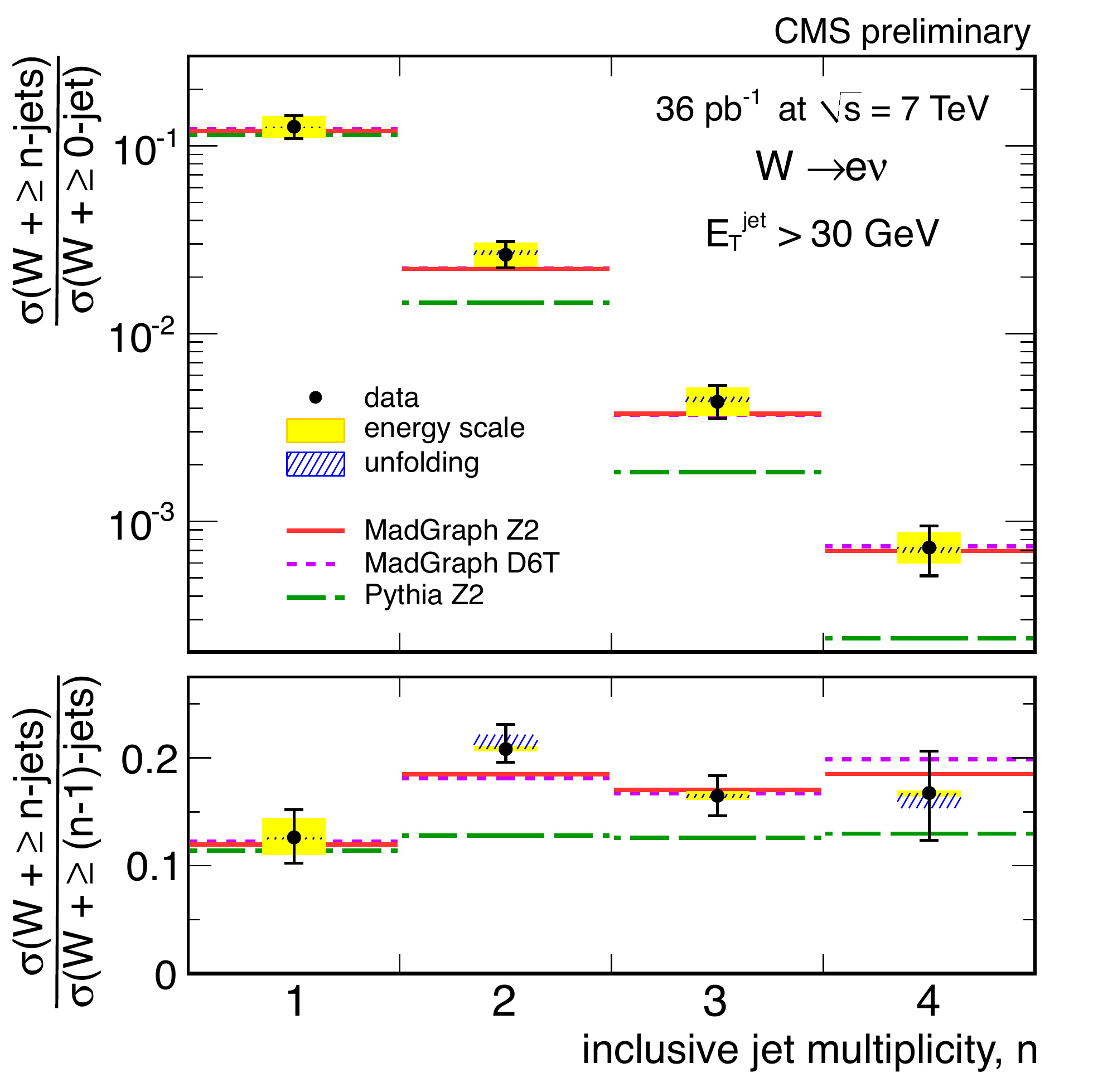}
  \includegraphics[width=0.4\textwidth]{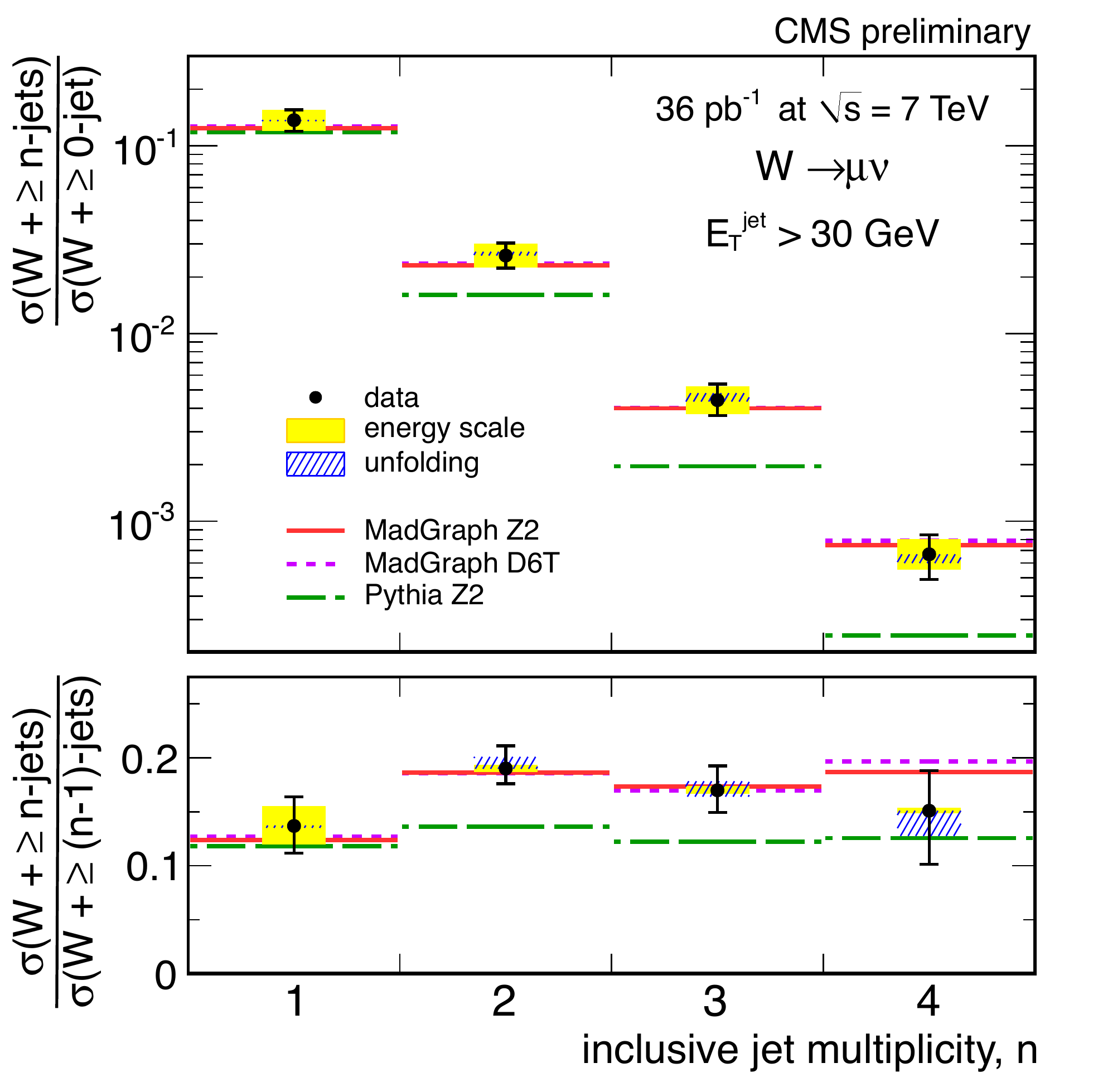}

  \caption{
The ratio  $\sigma(W+n~{\mathrm{jets}})/\sigma(W)$
in the electron channel (left) and muon channel (right)
compared to expectations from \MADGRAPH and \PYTHIA.
\label{fig:Wenurates}}
\end{figure}

\begin{figure}%[h!]
  \centering
  \includegraphics[width=0.4\textwidth]{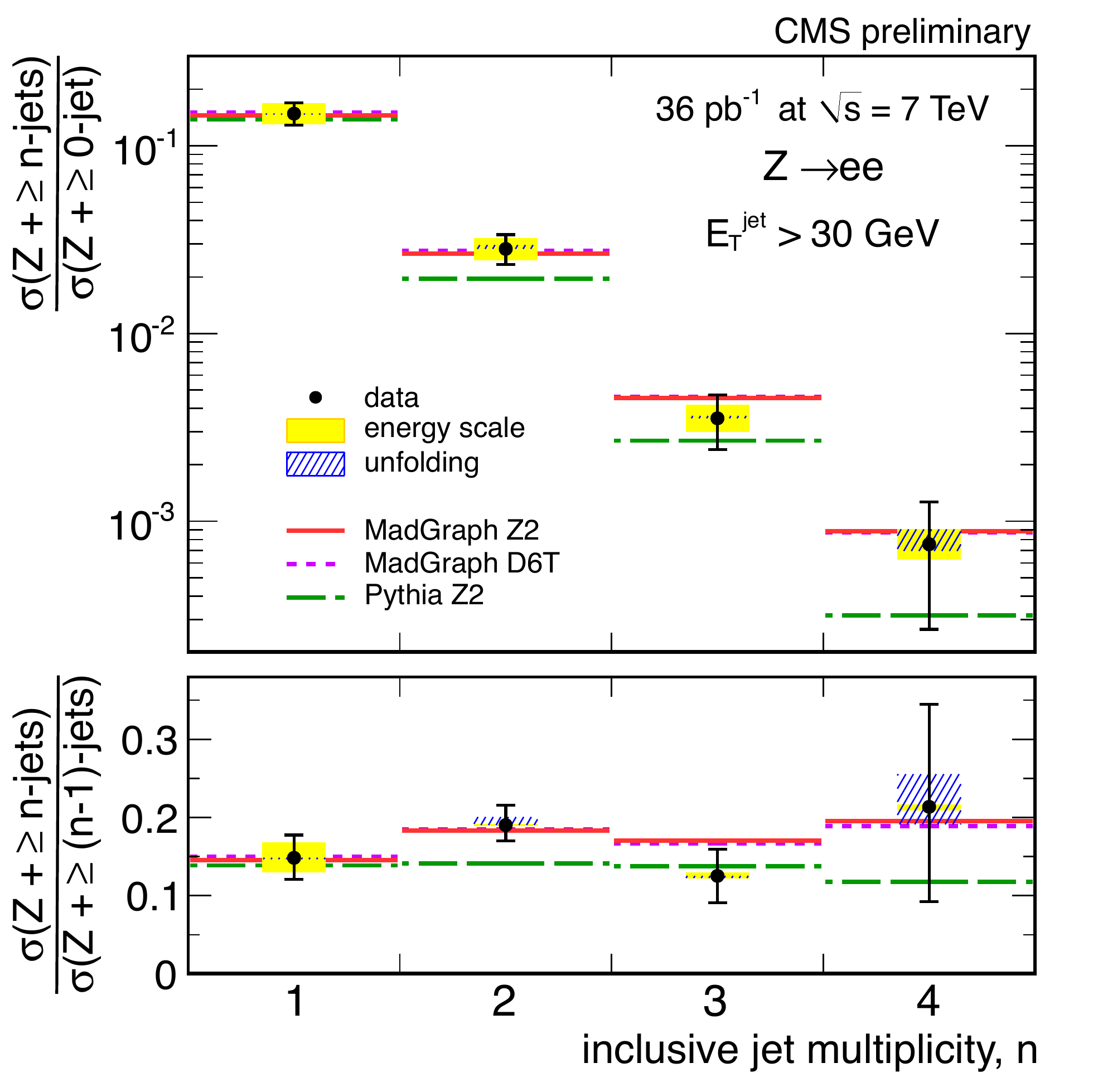}
  \includegraphics[width=0.4\textwidth]{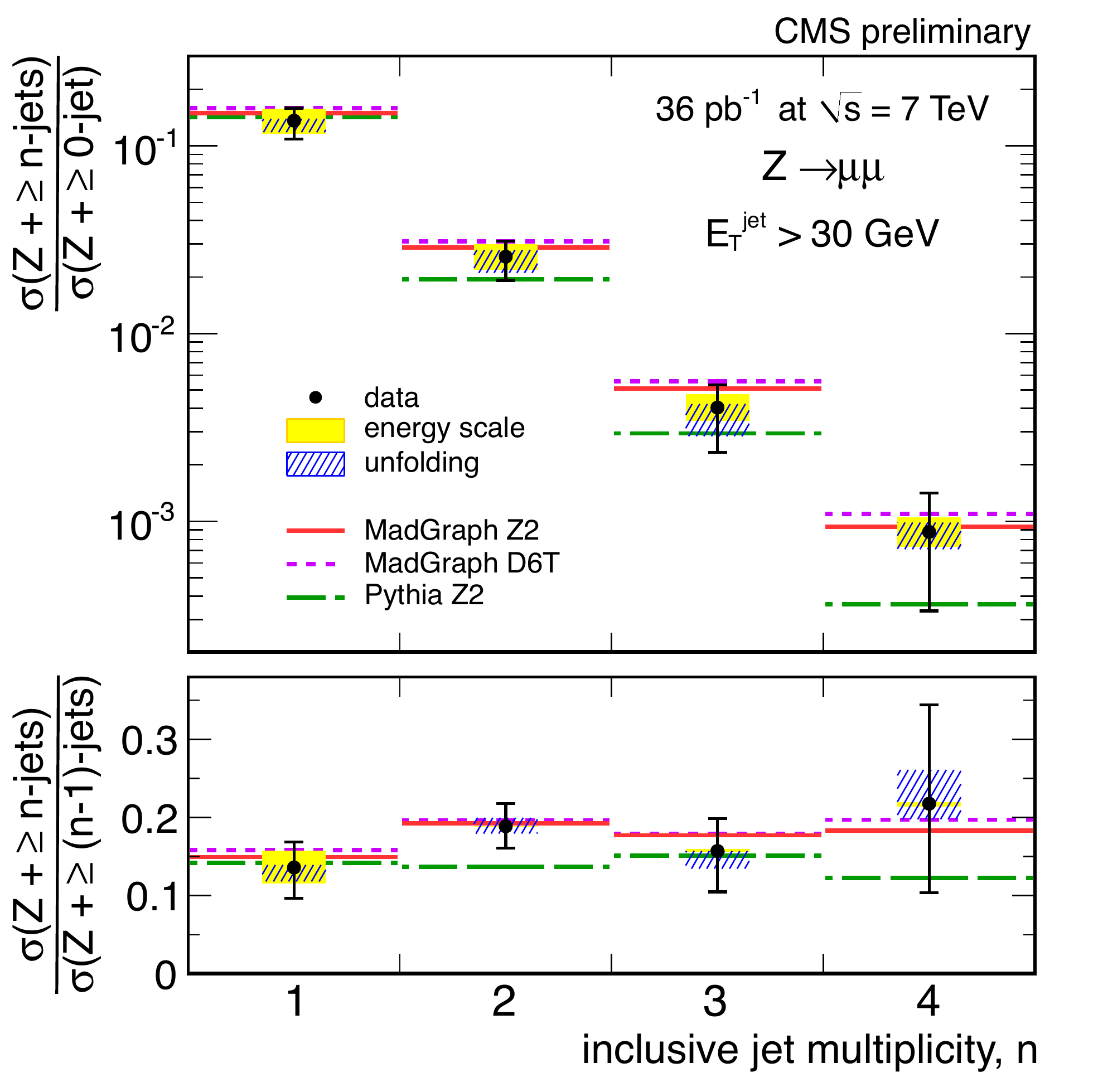}

  \caption{
The ratio  $\sigma(Z+n~{\mathrm{jets}})/\sigma(Z)$
in the electron (left) channel and muon channel (right)
compared to expectations from \MADGRAPH and \PYTHIA.
\label{fig:Zeerates}}
\end{figure}

From the unfolded exclusive jet multiplicity distributions we derive inclusive jet
multiplicities and calculate two sets of ratios.  The
first set of ratios is $\sigma(\PV+n~{\mathrm{jets}})/\sigma(\PV)$,
where $\sigma(\PV)$ is the inclusive cross section, see the upper frames of
Figs.~\ref{fig:Wenurates}--\ref{fig:Zeerates}.
The second set of ratios is 
$\sigma(\PV+n~{\mathrm{jets}})/\sigma(\PV+(n-1)~{\mathrm{jets}})$, shown in the lower frames of Figs.~\ref{fig:Wenurates}--\ref{fig:Zeerates}.
The systematic uncertainties associated with the JES
and the unfolding are shown as error bands. 
For a large number of jets, the \PYTHIA  
simulation fails to describe the data, while the
\MADGRAPH simulation agrees well, as expected.

\begin{figure}%[h!]
  \centering  
  {\includegraphics[width=0.8\textwidth]{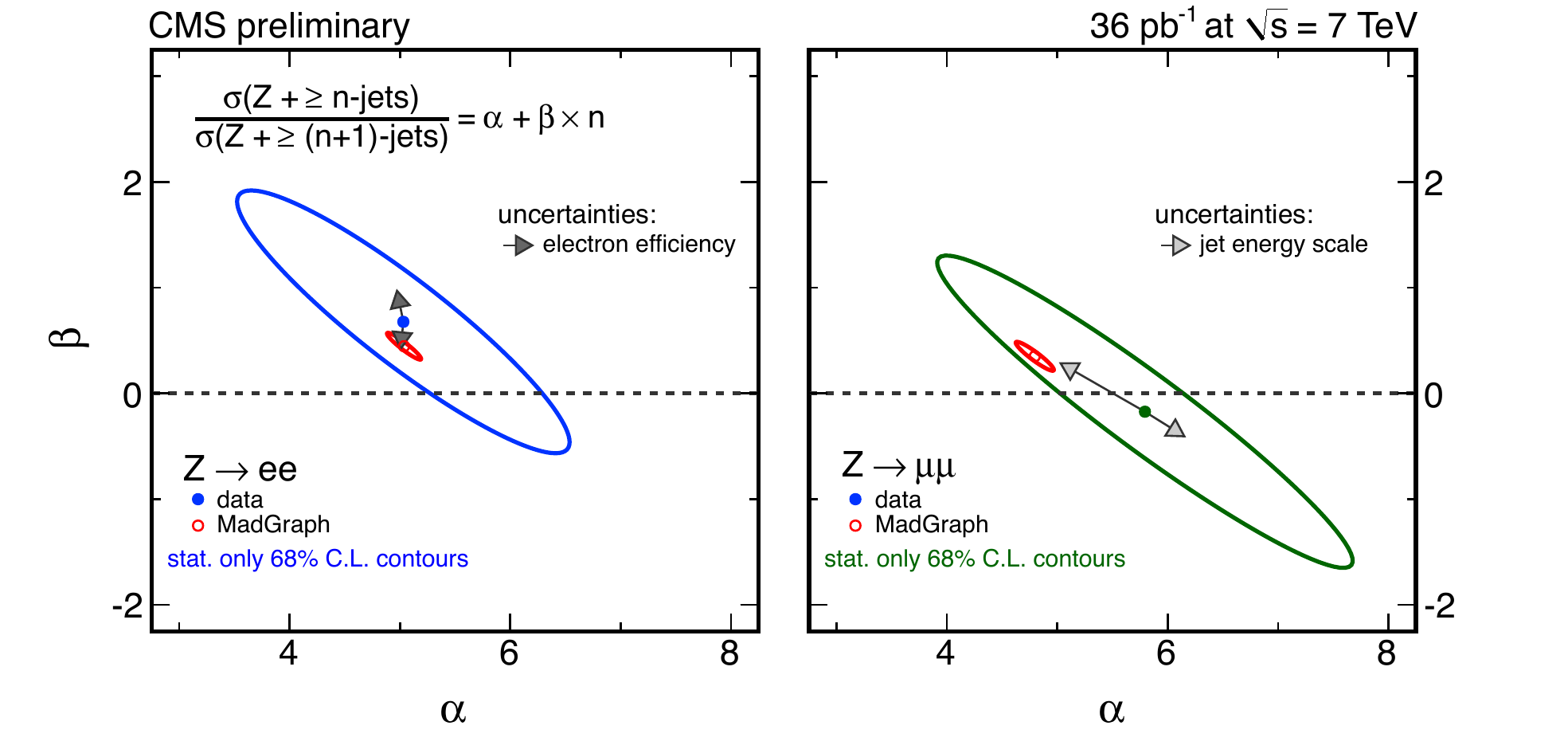}} \\
  {\includegraphics[width=0.8\textwidth]{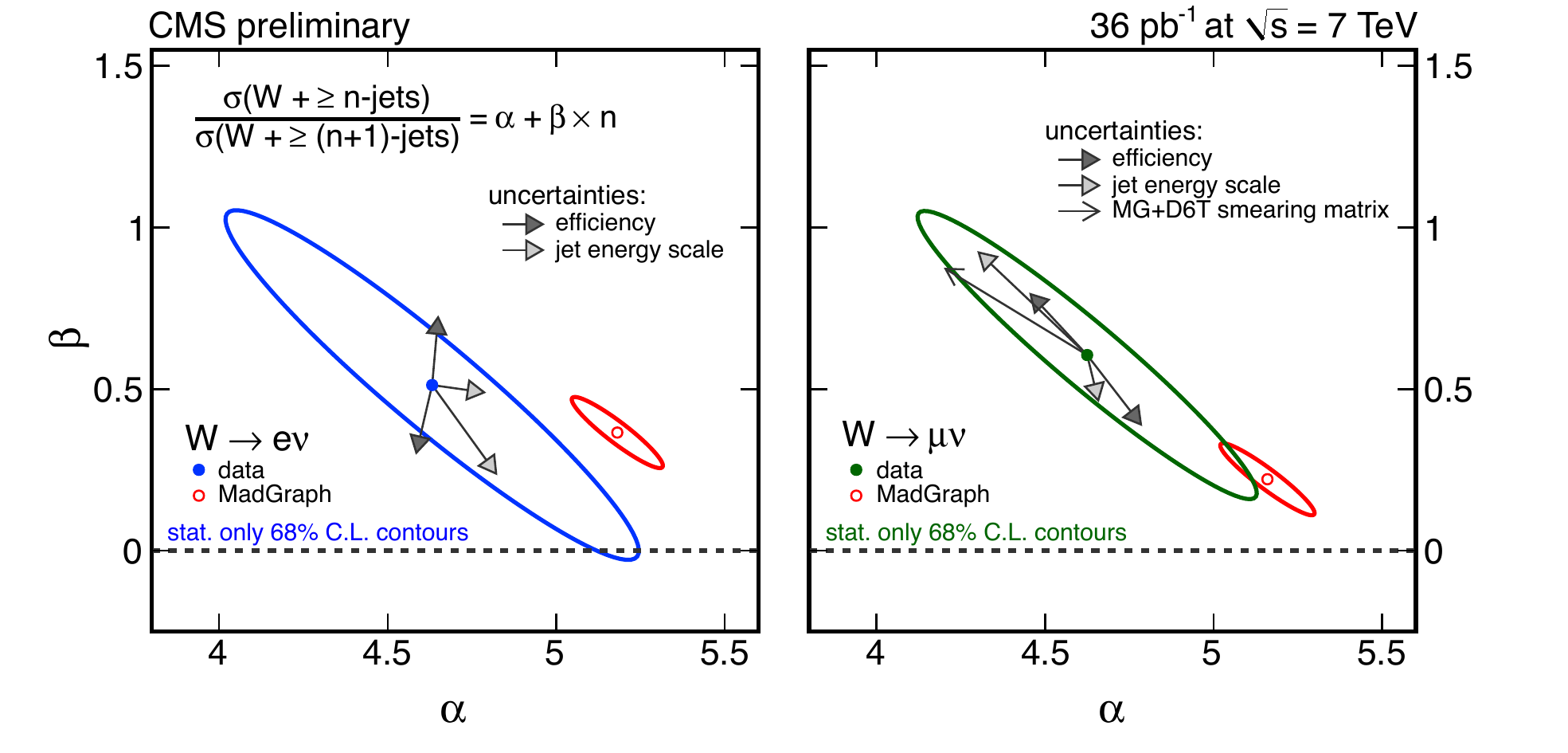}}
  \caption{Fit results on the Berends-Giele scaling parameters
  $\alpha$ and $\beta$ after pile--up subtraction, efficiency
  corrections, and unfolding for detector resolution effects. 
The data are compared to \MADGRAPH with the Z2 tune. 
a) shows \zjets, b) shows \wjets.
    \label{fig:alphabeta}
    }
\end{figure}

Finally, we show the results of the fit for $\alpha$ and $\beta$
in our treatment of Berends-Giele scaling for both 
\wjets and \zjetsb in Fig.~\ref{fig:alphabeta}.
The results are given in the $(\alpha,\beta)$
plane and are compared to the results obtained from the
\MADGRAPH sample.  The electron and muon expected
values differ mostly because of the $\Delta R>0.3$ cut between the jets and
the leptons, which is applied only in the electron channel. 
The ellipses correspond to 68\% C.L. considering only statistical uncertainty.
The arrows show the effect on the central value from 
the most important sources of systematic uncertainty.
The measurements agree well in the \zjetsb channels, and
fairly well in the \wjets channel.
The $\beta$ parameter is within one standard deviation from zero 
for the \wjets case and within 0.5 standard deviation for the \zjetsb. 
The values for \wjets and \zjetsb agree with one another,
as expected in the standard model. 
The data is found to be in reasonable agreement with the theoretical 
expectations with deviations that are within one or two standard deviations depending on the channel.

We measured the rate of jet production in association
with a \W or \Z vector boson using $pp$ collision data
at $\sqrt{s} = 7$~TeV.  The leading jet \PT spectrum agrees well with simulations
based on \MADGRAPH + \PYTHIA and the Z2 tune.
We unfolded the exclusive jet multiplicity distributions
and measured the ratios of cross sections
$\sigma(\PV+\ge n~{\mathrm{jets}})/\sigma(\PV)$ and 
$\sigma(\PV+\ge n~{\mathrm{jets}})/\sigma(\PV+\ge (n-1)~{\mathrm{jets}})$
where $n$ is the inclusive number of jets.  
The results are in agreement with the \MADGRAPH generator.
Finally, we made a quantitative test of Berends-Giele scaling.
The results show good agreement between \wjets and
\zjetsb and fair agreement with the simulation.

\begin{acknowledgments}
I would like to acknowledge all of the people who worked extremely hard on this analysis:
N. Akchurin, S.B.Beri, S. Bolognesi, A. Branca, Y. Chen, V. Ciulli, S. Dasu, B. Dahmes, J. Damgov, E. Di Marco, J. Z Efron, S. Frosali, E. Gallo, S. Gonzi, M. Grothe, P. Klabbers, S. Lacaprara, C. Lazaridis, S. W. Lee, P. Lenzi, J.Lykken, S. Malik, M. Nespolo, P. Meridiani, M. Mozer, M. Pierini, W. Reece, C. Rogan, I. Ross, C. Rovelli, L. K. Saini, A. Schizzi, I. Segoni, A. P. Singh, M. Spiropulu, P.Traczyk, L. Vanelderen, and E. Yazgan.

\end{acknowledgments}

\bigskip % extra skip inserted

\end{document}